\documentclass[12pt,preprint]{aastex}
\begin{document}
\author{Daniel J. Cross}
\affil{Department of Physics, Drexel University}
\affil{Philadelphia, PA 19104, USA}
\email{d.j.cross@drexel.edu}
\title{Comments on the Cooperstock-Tieu Galaxy Model}
\begin{abstract}
The recently proposed Cooperstock-Tieu galaxy model claims
to explain the flat rotation curves without dark matter.  The purpose
of this note is to show that this model is internally inconsistent and thus
cannot be considered a valid solution.  Moreover, by making the solution
consistent the ability to explain the flat rotation curves is lost.
\end{abstract}


\section{Introduction}
Cooperstock and Tieu model a galaxy in general relativity as an axially
symmetric, pressure free dust cloud with metric
\begin{equation} \label{metric}
ds^2=-e^{w}(cdt-Nd\varphi)^2+r^2e^{-w}d\varphi^2+e^{v-w}(dr^2+dz^2)
\end{equation}
where $w$, $v$, and $N$ are functions only of $r$ and $z$, thus the
metric is stationary \citep{CT}.  They further assume their coordinates to be
co-moving with the galactic dust, thus 
\begin{equation}
u^\mu=e^{-w/2}\delta_t^\mu
\end{equation}
is the four-velocity.

By performing a local diagonalization of the metric they obtain a
relation between the metric and local angular velocity as
\begin{equation}
\omega=\frac{cNe^w}{r^2e^{-w}-N^2e^w}\approx\frac{cN}{r^2}\label{rot}
\end{equation}
where the approximate value is appropriate to the order they consider.  The $tt$ and $t\varphi$ Einstein equations are respectively
\begin{mathletters}
\begin{eqnarray}
r^{-2}\left({N_{,r}^2+N_{,z}^2}\right) &=& \frac{8\pi G \rho}{c^2}\label{dens}\\ 
N_{,rr}-\frac{1}{r}N_{,r}+N_{,zz}&=&0. \label{neqn}
\end{eqnarray}
\end{mathletters}
The second equation is equivalent to the formula
\begin{equation} \label{lap}
\nabla^2\Phi=0
\end{equation}
where they have defined\footnote{It should be noted that this $\Phi$ is not
the Newtonian gravitational potential.}
\begin{equation}
\Phi=\int\frac{N}{r}dr.
\end{equation}

Thus the observed rotation curve becomes a boundary condition for
the solution to Laplace's equation (\ref{lap}) which they take in the form
\begin{equation}
\Phi=\sum_nC_ne^{-k_n|z|}J_0(k_nr)
\end{equation}
where the $k_n$ are chosen for orthogonality over the radius of the
galaxy.  Once $N$ is found by fitting this function to the
obsersved rotation curve, they derive the density by
(\ref{dens}) and in this way they obtain an excellent fit to the data
while obtaining a density profile that accords with observation.

However, it has been pointed out \citep{c1,c2} that, since the solution
depends on $|z|$, equation (\ref{neqn}) is not satisfied, but
rather yields a singular contribution to the $z=0$ plane, which has the
properties of an exotic form of matter.  It may be wondered whether
this singular disk can be removed by choosing a different solution
form or by increasing the complexity of the model.  However, in the
following analysis we will show that this is not possible and that
model fails to accord with general relativity.

\section{Analysis}
Assuming this form of the metric, and without making any
approximations, the scalar of volume expansion $\Theta \equiv
u^\mu_{\phantom{\mu};\mu}$ vanishes (a semicolon denotes covariant
differentiation and a comma denotes partial differentiation).  Defining the space-projection tensor
\begin{equation}
h_{\mu\nu}=g_{\mu\nu}+u_\mu u_\nu
\end{equation}
the shear tensor is given by\footnote{$(\mu,\nu)$
  and $[\mu,\nu]$ denote symmetrization and antisymmetrizaion with respect to
  the enclosed indices, respectively.}
\begin{equation}
\sigma_{\mu\nu}=u_{(\alpha;\beta)}h^\alpha_{\phantom{\alpha}\mu}h^\beta_{\phantom{\alpha}\nu}
\end{equation}
and this vanishes as well\footnote{The vanishing of shear was also
  found by Bonner in his solution \citep{B}.}.  However, the vorticity tensor, given by
\begin{equation}
\omega_{\mu\nu}=u_{[\alpha;\beta]}h^\alpha_{\phantom{\alpha}\mu}h^\beta_{\phantom{\alpha}\nu}
\end{equation}
does not vanish, but has nonzero components
\begin{mathletters}
\begin{eqnarray}
\omega_{\varphi r} = -\omega_{r\varphi}&=& \case{1}{2}e^{w/2}N_{,r}\\
\omega_{\varphi z} = -\omega_{z\varphi}&=& \case{1}{2}e^{w/2}N_{,z}.
\end{eqnarray}
\end{mathletters}
Though the matter in this model does indeed rotate, the rotation is
rigid and thus cannot characterize a galaxy which is differentially
rotating.  It should be emphasized that since $\sigma_{\mu\nu}$ is a
tensor, this cannot be a coordinate effect.

With the present solution Raychaudhuri's equation \citep{CW} simplifies to
\begin{equation}
-\omega_{\mu\nu}\omega^{\mu\nu}=R_{tt}/g_{tt}
\end{equation}
and in fact reduces to the condition $\nabla^2w=0$.  This condition is
demanded in \cite{CT} on the grounds that the geodesic equation be
satisfied.  This is equivalent to saying that the geodesics must be
circular orbits about the $z$-axis, which should not hold in general.
Orbits in the $z=0$ plane should indeed be azimuthal, but we cannot expect this
behavior off that plane.  That Raychaudhuri's equation demands this
condition again reveals the rigidity of the rotation\footnote{Actually,
any co-moving coordinate system requires $g_{tt}=-1$ as the coordinate
points are in free fall and thus keep proper time \citep{W}, which here
requires $w\equiv 0$.}.

Now, if we seek solutions to (\ref{neqn}) that are symmetric about the
plane, and singularity free, then must require $N$ to be independent\footnote{We could choose $\cosh(z)$, but this would lead to
  an exponentially increasing matter density.} of $z$.  Thus (\ref{neqn}) has
the trivial solutions
\begin{equation}
\begin{array}{ccc}
N = A & \textrm{or} & N=Br^2
\end{array}
\end{equation}
where $A$ and $B$ are constants.  The first solution leads to zero
density and the second to a constant density under rigid rotation,
according to equations (\ref{rot}) and (\ref{dens}).  Thus it appears
that the physical origin of the singularity is in attempting to
describe, in co-moving coordinates, a non-rigidly rotating dust cloud,
which the metric (\ref{metric}) cannot.

Next, in order to solve the Einstein Equations Cooperstock and Tieu perform an
expansion of the metric in  $\sqrt{G}$ and conclude that the functions
$w$ and $v$ are of second order, but the function $N$, which couples to the
rotation, is of first order.  Strictly speaking, this expansion is not
well defined as the expansion parameter has dimensions.  We can form
the dimensionless parameter 
\begin{equation} \label{lambda}
\lambda=\sqrt{\frac{GM}{Lc^2}}
\end{equation}
where $M$ is some characteristic mass of the system and $L$ some
characteristic length (for example, the mass and radius of the
galactic core).  We then compare equations (\ref{neqn}) and (\ref{rot}):
\begin{eqnarray*}
\lambda^2r^{-2}\left(N_{,r}^2+N_{,z}^2\right) &=& \frac{8\pi
  G\rho}{c^2} \\
\lambda\frac{Nc}{r} &=& v 
\end{eqnarray*}
where the order has been shown explicitly.  Substituting $v$ for $N$
in the first equation yields the relation
\begin{equation}
v=\mathcal{O}\left(\sqrt{8\pi G \rho L^2}\right)
\end{equation}
where we have taken derivatives to be of order $1/L$.  This can be
rewritten as 
\begin{equation}
v=\mathcal{O}\left(c\lambda \sqrt{\frac{\rho L^3}{M}}\right)
\end{equation}
and since the quantity under the square root is of order unity, we
have
\begin{equation}
\frac{v}{c}=\mathcal{O}(\lambda)=\mathcal{O}\left(\sqrt{\frac{GM}{Lc^2}}\right)\label{order}
\end{equation}
which is expected from Newtonian theory and is the basis of the PPN
expansion.

Now, suppose we choose a coordinate system in which the galactic dust
has coordinate velocity $\omega/c$, so that the stress-energy tensor has
the form
\begin{equation}
T^{\mu\nu} \propto \left( \begin{array}{cc}
1 & \displaystyle \frac{\omega}{c} \\
\\
\displaystyle \frac{\omega}{c} &  \displaystyle \left(\frac{\omega}{c}\right)^2
\end{array} \right)
\end{equation}
in the $t\varphi$-subspace.  The $t\varphi$-Einstein equation then has the
form
\begin{equation}
G^{t\varphi}=\frac{8\pi G}{c^2}T^{t\varphi}\propto \frac{8\pi
  G}{c^2}\frac{\omega}{c}
\end{equation}
up to a factor of order unity due to the constraint condition $u_\mu
u^\mu=-1$.  Thus, the right hand side of this equation begins at order
$\lambda^3/L^2$ according to (\ref{order}), whereas the left hand
side is of order $\lambda/L^2$ according to (\ref{neqn}), since $N$ is assumed to be of order $\lambda$.  Thus the assumption that $N$ is of
first order is inconsistent, while consistency requires that $N$ be of at
least third order\footnote{This will be demonstrated more explicitly in the next section.}.

Moreover, given the form of the stress-energy tensor above, suppose we
make a \emph{global} transformation $\varphi \to \varphi+\omega(r,z)t$
to the co-moving frame, so that all components of the new
stress-energy tensor vanish except the density. The new metric will
have the same form as the old metric, but for the differential $d\varphi'$ we have
\begin{equation}
d\varphi'=\omega dt + d\varphi +t(\omega_{,r}dr+\omega_{,z}dz)
\end{equation}
which necessarily introduces time-dependence into the new metric
unless $\omega$ is spatially-independent, that is, unless the rotation
of the matter is rigid.  Thus, contrary to the assumption of
Coopertock and Tieu, the metric (\ref{metric}) cannot both be
co-moving and time-independent.  This accords with zero value of the shear tensor above.

It can be seen in the following figure that a co-moving metric of a
differentially rotating system is time-dependent and possesses shear.  Here,
$r$-coordinate lines ``twist'' up in time relative to observers at
spatial infinity.  $\varphi=\omega t$ has been plotted, where
$\omega(r)$ is the fit to the Milky Way from \cite{CT}.
\begin{figure}[h]
\begin{center}
\epsscale{.5}
\plottwo{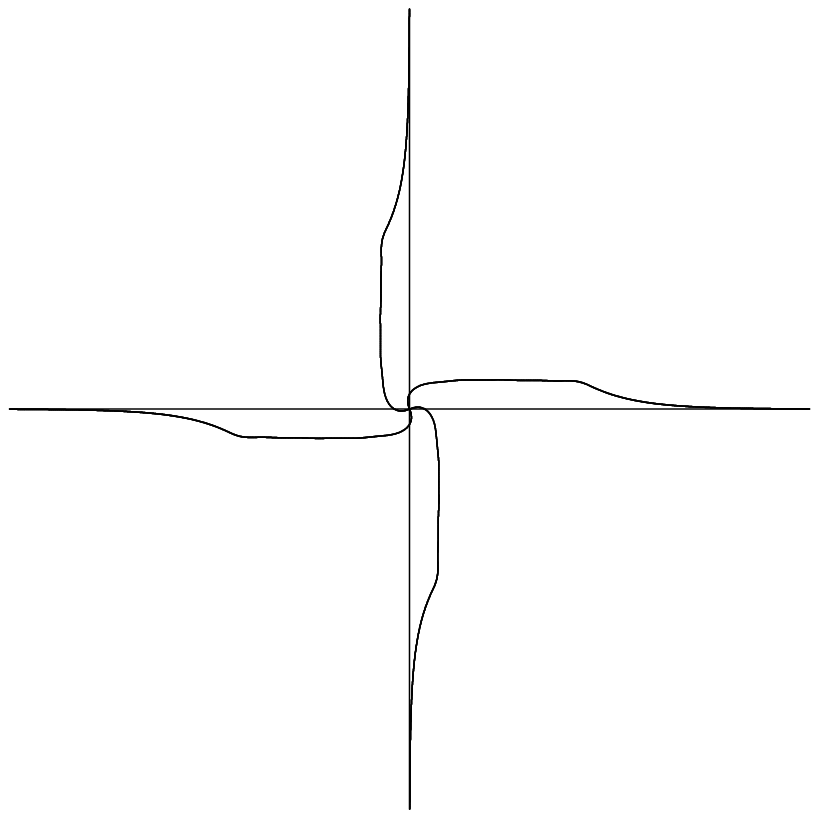}{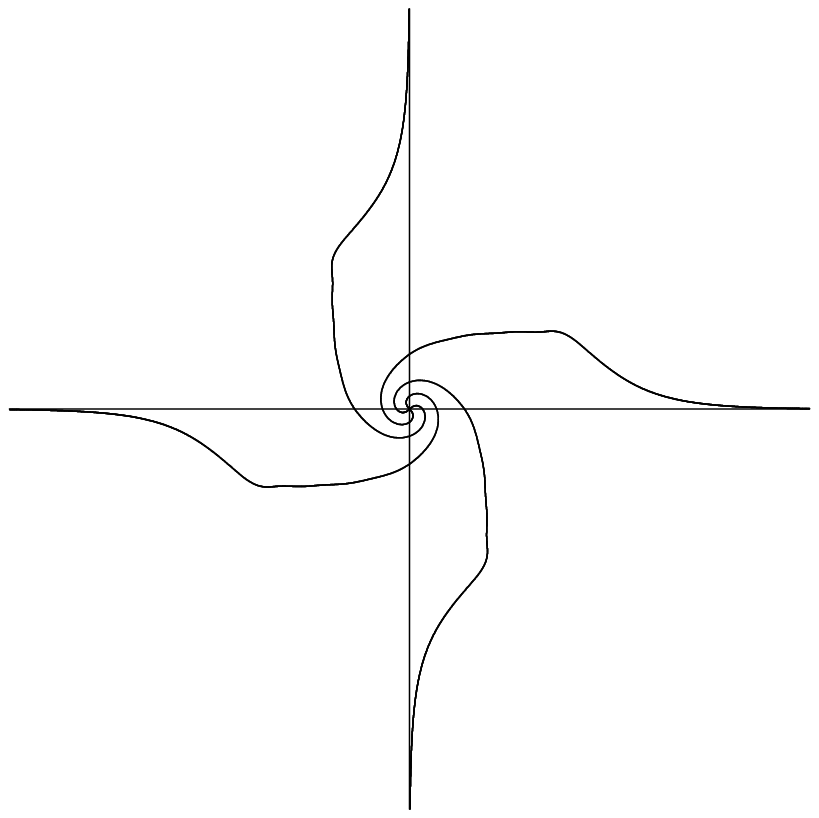}
\caption{The twisting of $r$-coordinate lines for various $\varphi$ at
two different times as seen by observers at spatial infinity.}\label{fig1}
\end{center}
\end{figure}

\section{Non-co-moving Expansion}
In this section we carry out an expansion of the metric
(\ref{metric}) in a system of reference in which the galactic
dust has coordinate velocity 
\begin{equation}
\frac{u^2}{u^0}=\frac{d\varphi}{dct}=\frac{\omega}{c}
\end{equation}
which is physically the angular velocity measured by observers at
spatial infinity \citep{BW}.  In this direct approach we will show that
the angular momentum coupling is too weak to account
for the flat rotation curves.

We expand the metric (\ref{metric}) as
\begin{mathletters}
\begin{eqnarray}
g_{tt} &=& -1 - \stackrel{2}{w}-\stackrel{4}{w}+\mathcal{O}(\lambda^6) \\
g_{t\varphi} &=&
\stackrel{1}{N}+\stackrel{2}{w}\stackrel{1}{N}+\stackrel{3}{N} +\mathcal{O}(\lambda^5) \\
g_{\varphi\varphi} &=& r^2 -\stackrel{1}{N^2}-r^2\stackrel{2}{w}+\mathcal{O}(\lambda^4) \\
g_{rr} =g_{zz} &=& 1 +\stackrel{2}{w}-\stackrel{2}{v}+\mathcal{O}(\lambda^4)
\end{eqnarray}
\end{mathletters}
where the over-script indicates the order of the term in $\lambda$,
which is the same as defined in (\ref{lambda}). The presence of
only even terms in the $g_{\mu\mu}$ and odd in $g_{t\varphi}$ are for the
appropriate behavior under time-reversal \citep{W}.

The constraint condition $u_{\mu}u^{\mu}=-1$ requires that 
\begin{equation}
u^t=(g_{tt}+2\frac{\omega}{c}g_{t\varphi}+\frac{\omega^2}{c^2}g_{\varphi\varphi})^{-1/2}
\end{equation}
and thus the stress energy tensor has the expansion
\begin{mathletters}
\begin{eqnarray}
T^{tt} &=& \rho{c^2}\left( 1- \stackrel{2}{w} +\frac{r^2\omega^2}{c^2} \right) +\mathcal{O}(\lambda^4)\\
T^{t\varphi} &=& \rho{c^2}\frac{\omega}{c} +\mathcal{O}(\lambda^3)\\
T^{\varphi\varphi} &=& \rho{c^2}\frac{\omega^2}{c^2}+\mathcal{O}(\lambda^4)
\end{eqnarray}
\end{mathletters}
As seen in the previous section, $8\pi G/c^2$ increases the order by
two so that the right-hand side of the Einstein equations are of second order and higher, thus the $t\varphi$ equation to first order is
\begin{equation}
\stackrel{1}{N}_{,rr}-\frac{1}{r}\stackrel{1}{N}_{,r}+\stackrel{1}{N}_{,zz}=0
\end{equation}
so that the lowest order term of $N$ is sourceless.  We are free to
choose $\stackrel{1}{N}$ however we wish to make the solution simplest, thus we set
$\stackrel{1}{N}=0$.  With this selection the Einstein equations
through third order become
\begin{mathletters}
\begin{eqnarray}
-\nabla^2\stackrel{2}{w}+\case{1}{2}\left( \stackrel{2}{v}_{,zz}+
\stackrel{2}{v}_{,rr} \right) &=& \frac{8\pi G \rho}{c^2}\\
\stackrel{2}{v}_{,r}&=& 0\\
\stackrel{2}{v}_{,zz}+\stackrel{2}{v}_{,rr}&=& 0\\
\stackrel{2}{v}_{,z} &=& 0\\
-\case{1}{2}r^{-2}\left(
\stackrel{3}{N}_{,rr}-\frac{1}{r}\stackrel{3}{N}_{,r}+\stackrel{3}{N}_{,zz}
\right) &=& \frac{8\pi G \rho r\omega}{c^3}.
\end{eqnarray}
\end{mathletters}
We see that  $\stackrel{2}{v}$ must be a constant so that we have
\begin{mathletters}
\begin{eqnarray}
\nabla^2\stackrel{2}{w} &=& -\frac{8\pi G \rho}{c^2}\\
\stackrel{3}{N}_{,rr}-\frac{1}{r}\stackrel{3}{N}_{,r}+\stackrel{3}{N}_{,zz}
&=& -\frac{16\pi G \rho r^3\omega}{c^3}.
\end{eqnarray}
\end{mathletters}
Thus we see that the coupling to the angular momentum is of third
order, there is no longer a nonlinear term in the mass
density\footnote{Even when $G^{tt}$ is written to fourth order the
  nonlinearity due to $N$ is not present.}, and we can identify 
\begin{equation}
\stackrel{2}{w}=-\frac{2\Phi}{c^2}
\end{equation}
with the Newtonian gravitational potential.

If we analyze circular orbits on the plane the geodesic equation
demands that
\begin{equation}
\Gamma^\mu_{tt}+2\Gamma^\mu_{t\varphi}\frac{\omega}{c}+ \Gamma^\mu_{\varphi\varphi}\frac{\omega^2}{c^2}=0
\end{equation}
which to third order can be written for $\mu=r$ as
\begin{equation}
\frac{v^2}{r}=-\Phi_{,r}\left( 1+\frac{v^2}{c^2}\right)-\frac{vc}{r}\stackrel{3}{N}_{,r}
\end{equation}
which is recognized as the Newtonian centripetal equation plus second
order corrections\footnote{The presence of the $c$ in the last term
  effectively lowers the order by one.}.  Thus the matter essentially
moves according to the predictions of Newtonian gravitation with
corrections of order $v^2/c^2$, which cannot account for flattening of
the rotation curves without extra non-luminous matter.

Finally, we can compute the shear tensor, which has the non-zero
components
\begin{mathletters}
\begin{eqnarray}
\sigma_{tr}=\sigma_{rt} &=& -\frac{(u^t)^3r^2}{2c^2}\omega\omega_{,r}\\
\sigma_{tz}=\sigma_{zt} &=& -\frac{(u^t)^3r^2}{2c^2}\omega\omega_{,z}\\
\sigma_{\varphi r}=\sigma_{r\varphi} &=& \frac{(u^t)^3r^2}{2c}\omega_{,r}\\
\sigma_{\varphi z}=\sigma_{z\varphi} &=& \frac{(u^t)^3r^2}{2c}\omega_{,r}
\end{eqnarray}
\end{mathletters}
all of which vanish exactly when $\omega$ is constant.

\section{Conclusion}
It has been shown that the Cooperstock-Tieu galaxy model is
inconsistent as a general relativistic model and that a proper model
fails to account for the flatness of the rotation curves without the
dark-matter hypothesis.  This failure is due to the weakness of the
metric coupling to the angular momentum of the galaxy.  

However, the flat rotation curves seem to imply a large inertial induction
effect, where the rotating inner matter boosts the rotation of the outer
matter, leveling off the rotation curve, which is what the
Cooperstock-Tieu model attempts to describe within general
relativity.  Since their solution predicts a matter density well
within visible limits it is quite possible that their solution
represents an alternative, more Machian, gravitational theory where
inertial induction effects are much larger than in General Relativity.

\acknowledgements

The author is greatful to Prof. R. Gilmore for his encouragement and
many helpful discussions on this project.

\end{document}